\documentclass[12pt]{amsart}
\usepackage{amssymb}
\usepackage{fullpage}
\usepackage{graphicx}
\usepackage{enumerate}
\input pictex.sty

\newtheorem{thm}{Theorem}[section]

\newtheorem{lem}[thm]{Lemma}

\theoremstyle{definition}

\theoremstyle{remark}

\numberwithin{equation}{section}

\newcommand{\pd}{\partial}

\newcounter{space}

\begin{document}

\title{Self-Similar Solutions of the Non-Strictly Hyperbolic Whitham Equations}%
\author{V. U. Pierce}
\address{Department of Mathematics, Ohio State University, 231 W. 18th Avenue, Columbus, OH 43210}%
\email{vpierce@math.ohio-state.edu}
\author{Fei-Ran Tian}%
\address{Department of Mathematics, Ohio State University, 231 W. 18th Avenue, Columbus, OH 43210}
\email{tian@math.ohio-state.edu}

\thanks{}%

\begin{abstract}
We study the Whitham equations for the
fifth order KdV equation. The equations are neither strictly hyperbolic nor
genuinely nonlinear. We are interested in the solution of the Whitham equations
when the initial values are given by a step function. We classify the step like initial data 
into eight different
types. We construct self-similar solutions for each type.
\end{abstract}
\maketitle

\markboth{ V.U. PIERCE AND F.-R. TIAN}{
SELF-SIMILAR SOLUTION OF THE WHITHAM EQUATIONS}
\pagestyle{myheadings}

\section{Introduction}

It is known that the solution of the KdV equation
\begin{equation} \label{eq1}
u_t + 6 u u_x + \epsilon^2 u_{xxx}  =  0 
\end{equation} 
has a weak limit as $\epsilon \to 0$ while the initial values 
$$u(x, 0; \epsilon) = u_0(x)$$
are fixed.
This weak limit satisfies the Burgers equation
\begin{equation}
\label{Burgers}
u_t + (3 u^2)_x = 0
\end{equation}
until its solution develops shocks. Immediately after, the weak limit is governed
by the Whitham equations \cite{lax, lax2, ven, whi} 
\begin{equation}
\label{KdVW}
u_{it} + \lambda_i(u_1, u_2, u_3) u_{ix} = 0 \ , \quad i=1, 2, 3,
\end{equation}
where the $\lambda_i$'s are given by formulae (\ref{lambda}).
After the breaking of the solution of (\ref{KdVW}), the weak limit is described by systems of
at least five hyperbolic equations similar to (\ref{KdVW}).

The KdV equation (\ref{eq1}) is just the first member of an infinite sequence
of equations, the second of which is the so-called fifth order KdV equation
\begin{equation}
\label{5KdV}
u_t + 30 u^2 u_x + 20 \epsilon^2 u_x u_{xx} + 10 \epsilon^2 u u_{xxx} + \epsilon^4 u_{xxxxx} = 0 \ .
\end{equation}
The solution of the fifth order KdV equation (\ref{5KdV}) also has a weak limit as $\epsilon \to 0$.
As in the KdV case, this weak limit satisfies the Burgers type
equation 
\begin{equation}
\label{5Burgers}
u_t + (10 u^3)_x = 0 
\end{equation}
until the solution of (\ref{5Burgers}) forms a shock. Later, the limit is governed by equations similar
to (\ref{KdVW}), namely,
\begin{equation}
\label{5KdVW}
u_{it} + \mu_i(u_1, u_2, u_3) u_{ix} = 0 \ , \quad i=1, 2, 3,
\end{equation}
where $\mu_i$'s are given in (\ref{eq18a}). They will be also be called the Whitham equations.

In this paper, we are interested in the solution of the Whitham equation (\ref{5KdVW}) 
for the fifth order KdV (\ref{5KdV}) with a step like initial function
\begin{equation} \label{step}
u_0(x) = \left\{ \begin{matrix} a & x < 0 \\
b & x > 0 \end{matrix} \right. \ , \ \quad a \neq b \ .
\end{equation}
For such an initial function with $a>0$, $b<a$ or $a< 0$, $b>a$, the solution of the Burgers type 
equation (\ref{5Burgers}) has
already developed a shock at the initial time, $t=0$.
Hence, immediately
after $t=0$, the Whitham equations (\ref{5KdVW}) kick in.  Solutions of
(\ref{5KdVW}) occupy some domains of the space-time while solutions of
(\ref{5Burgers}) occupy other domains.  These solutions are matched on
the boundaries of the domains.

The solution of the Burgers equation (\ref{Burgers}) with initial function (\ref{step}) is simple: it is
either a rarefaction wave or a single shock wave. 

The Burgers type equation (\ref{5Burgers}) is more complicated, as its flux function
changes convexity at $u=0$. Its solution with step like initial data (\ref{step}) can
be a rarefaction wave, a single shock wave or a combination of both \cite{lef}. As a consequence,
the solutions of the Whitham equations (\ref{5KdVW}) will be seen to be more complex than those
of (\ref{KdVW}) in the KdV case. Indeed, there are eight types of different solutions in the former case
while there is only one type of solution in the latter case.

The KdV case with the step like initial data (\ref{step}) was first studied by Gurevich and Pitaevskii
\cite{gur}.  The Burgers solution of (\ref{Burgers}) develops a shock only for $a > b$. Moreover, the 
corresponding initial function is equivalent to the case $a=1$, $b=0$. In this case, Gurevich and Pitaevskii
found a self-similar solution of the Whitham
equations (\ref{KdVW}).  Namely, the space-time is divided into three parts 
$$(1) \   \frac{x}{t} < -6 \ , \quad
(2) \   -6 < \frac{x}{t} < 4  \ , \quad
(3) \  \frac{x}{t} > 4 \ .$$ 
The solution of (\ref{Burgers}) occupies the first and third parts, 
\begin{equation} 
\label{eq9a}
u(x, t) \equiv 1  \quad \mbox{when $\frac{x}{t} < -6$} \ , \quad 
u(x, t) \equiv 0 \quad  \mbox{when $\frac{x}{t} > 4$} \ .
\end{equation}
The Whitham solution of (\ref{KdVW}) lives in the second part, 
\begin{equation}
u_1(x, t) \equiv 1 \ , \quad
\frac{x}{t} = \lambda_2(1, u_2, 0) \ , \quad
u_3(x, t) \equiv 0 \ , 
\label{eq9b}
\end{equation}
when $-6 < x/t < 4$.  

Whether the second equation of (\ref{eq9b}) can be inverted to give $u_2$ as a function of the
self-similarity variable $x/t$ hinges on whether 
\begin{equation*}  
\frac{\pd \lambda_2}{\pd u_2} (1, u_2, 0) \neq 0. 
\end{equation*}
Indeed, Levermore \cite{lev} has proved the genuine nonlinearity of
the Whitham equations (\ref{KdVW}), i.e., 
\begin{equation}\label{eq11}
\frac{\pd \lambda_i}{\pd u_i} (u_1, u_2, u_3) > 0, \,
\quad i=1, 2, 3, 
\end{equation}
for $u_1 > u_2 > u_3$.  

For the fifth order KdV (\ref{5KdV}),  equations (\ref{5KdVW}), in
general, are not genuinely nonlinear, i.e., a property like
(\ref{eq11}) is not available.  Hence, solutions like (\ref{eq9a}) and (\ref{eq9b})
need to be modified.  

Our construction of solutions of the Whitham
equation (\ref{5KdVW}) makes use of the non-strict
hyperbolicity of the equations.  For KdV, it is known that the Whitham
equations (\ref{KdVW}) are strictly hyperbolic, namely:
\begin{equation*}
\lambda_1(u_1, u_2, u_3) > \lambda_2(u_1, u_2, u_3) > 
\lambda_3(u_1, u_2, u_3) 
\end{equation*}
for $u_1 > u_2 > u_3$.  For the fifth order KdV (\ref{5KdV}), different eigenspeeds of (\ref{5KdVW}),
$\mu_i(u_1, u_2, u_3)$'s, may coalesce in the region $u_1 > u_2 > u_3$.  

For the fifth KdV with step-like initial function (\ref{step}) where $a=1$ and $b=0$, the space time is 
divided into four regions (see Figure 1.) 
$$(1) \ \frac{x}{t} < - 15 \ , \quad
(2) \ -15 < \frac{x}{t} < \alpha \ , \quad
(3) \ \alpha < \frac{x}{t} < 16 \ , \quad
(4) \ \frac{x}{t} > 16 \ ,$$
where $\alpha$ is determined by (\ref{alpha}).
In the first and fourth regions, the solution of (\ref{5Burgers}) governs
the evolution:
$$u(x, t) \equiv 1 \quad \mbox{where $x/t < - 15$} \  \mbox{and} \ 
u(x, t) \equiv 0 \quad \mbox{where $x/t > 16$} \ .$$
The Whitham solution of (\ref{5KdVW}) lives in the second and third
regions; namely:
\begin{equation} 
\label{ns}
u_1(x, t) \equiv 1 \ , \quad 
\frac{x}{t} = \mu_2(1, u_2, u_3) \ , \quad
\frac{x}{t} = \mu_3(1, u_2, u_3) \ , 
\end{equation}
when $-15 < x/t < \alpha$, and
\begin{displaymath}
u_1(x, t) \equiv 1 \ , \quad 
\frac{x}{t} = \mu_2(1, u_2, 0) \ , \quad  u_3(x, t) \equiv 0 \ , 
\end{displaymath}
when $\alpha < x/t < 16$.

Equations (\ref{ns}) yield
$$ \mu_2(1, u_2, u_3) = \mu_3(1, u_2, u_3)$$
on a curve in the region $0< u_3 < u_2 < 1$.
This implies the non-strict hyperbolicity of the Whitham equations (\ref{5KdVW}) for
the fifth order KdV.

\begin{figure}[h] \label{fig1}
\begin{center}
\resizebox{12cm}{5cm}{\includegraphics{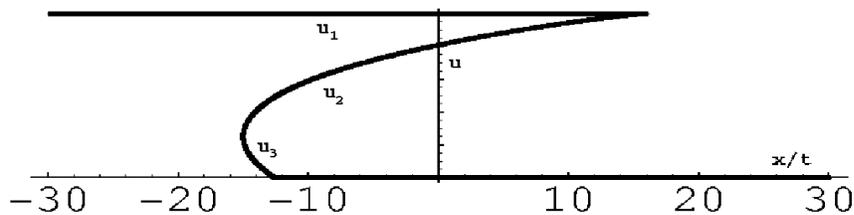}}
\caption{Self-Similar solution of the Whitham equations for
  $a=1$ and $b=0$ of type II.}
\end{center}\end{figure} 

The organization of the paper is as follows. In Section 2, we will study the
eigenspeeds, $\mu_i(u_1,u_2,u_3)$'s, of the Whitham equations (\ref{5KdVW}).
In Section 3, we will construct the self-similar solution of the Whitham equations
for the initial function (\ref{step}) with $a=1$, $b=0$.
In Section 4, we will use the self-similar solution of Section 3 to
construct the minimizer of a variational problem for the zero dispersion limit 
of the fifth order KdV. In Section 5, we will consider all the other possible step like
initial data (\ref{step}). We find that there are eight different types of
initial data. We construct self-similar solutions for each type.


\section{The Whitham Equations}

In this section we define the eigenspeeds of the Whitham equations for
both the KdV (\ref{eq1}) and fifth order KdV (\ref{5KdV}). We first introduce the 
polynomials of $\xi$ for $n=0, 1,
2, \dots$ \cite{dub, kri, Tian3}:
\begin{equation} \label{eq15} 
P_n(\xi, u_1, u_2, u_3) = \xi^{n+1} + a_{n, 1} \xi^n + \dots + a_{n,
  n+1} \ ,
\end{equation}
where the coefficients, $a_{n, 1}, a_{n, 2}, \dots, a_{n, n+1}$ are
uniquely determined by the two conditions
\begin{equation} \label{eq16}
\frac{ P_n(\xi, u_1, u_2, u_3) }{\sqrt{ (\xi - u_1)(\xi-u_2)(\xi-u_3)}
} = \xi^{n-1/2} + \mathcal{O}(\xi^{-3/2}) \quad  \mbox{for large $|\xi|$} 
\end{equation}
and
\begin{equation} \label{eq17}
\int_{u_3}^{u_2} \frac{ P_n(\xi, u_1, u_2, u_3)}{\sqrt{ (\xi- u_1)
    (\xi-u_2)(\xi-u_3)}} d\xi = 0 \ .
\end{equation}
Here the sign of the square root is given by $\sqrt{1}=1$.

In particular, 
\begin{equation}
\label{P01}
P_0(\xi,u_1, u_2, u_3) = \xi + a_{0, 1} \ , \quad
P_1(\xi,u_1, u_2, u_3) = \xi^2 - \frac{1}{2} (u_1 + u_2 + u_3) \xi + a_{1, 2} \ ,
\end{equation}
where 
\begin{align*}
a_{0, 1} &= (u_1 - u_3) \frac{E(s)}{K(s)} - u_1 \ , \\
a_{1, 2} &= \frac{1}{3} ( u_1 u_2 + u_1 u_3 + u_2 u_3 ) + \frac{1}{6}
(u_1 + u_2 + u_3) a_{0, 1} \ . 
\end{align*}
Here 
\begin{equation*}
s = \frac{u_2 - u_3}{u_1 - u_3} 
\end{equation*}
and $K(s)$ and $E(s)$ are complete elliptic integrals of the first and
second kind.  

$K(s)$ and $E(s)$ have some well-known properties \cite{Tian1, Tian2}. As $-1$ $<$ s $<$ $1$,
we have
\begin{eqnarray}
K(s) & = & \frac{\pi}{2} [1 + \frac{s}{4} + \frac{9}{64} s^{2}
+ \cdots + (\frac{1 \cdot 3 \cdots (2n-1)}{2 \cdot 4 \cdots 2n})^{2} s^{n}
+ \cdots] \ , \label{K}\\
E(s) & = & \frac{\pi}{2} [1 - \frac{s}{4} - \frac{3}{64} s^{2}
 - \cdots - \frac{1}{2n-1}(\frac{1 \cdot 3 \cdots (2n-1)}{
2 \cdot 4 \cdots 2n})^{2} s^{n} -
\cdots] \ , \label{E}
\end{eqnarray}
while, as $1 - s$ $\ll$ $1$, we have
\begin{eqnarray}
K(s) & \approx & \frac{1}{2} \log \frac{16}{1 - s}  \label{K2} \ , \\
E(s) & \approx & 1 + \frac{1}{4}(1 - s)[\log \frac{16}{1 - s} - 1] \ . \label{E2} 
\end{eqnarray}
Furthermore,
\begin{eqnarray}
\frac{d K(s)}{d s} & = & \frac{E(s) - (1-s)K(s)}{2s(1-s)} \ , \label{K3} \\
\frac{d E(s)}{d s} & = & \frac{E(s) - K(s)}{2s} \ . \label{E3} 
\end{eqnarray}
It immediately follows from (\ref{K}) and (\ref{E}) that
\begin{equation}
\frac{1}{1 - \frac{s}{2}} < \frac{K(s)}{E(s)} < \frac{1-\frac{s}{2}}{1-s}
\hspace*{.5in} for ~ 0 < s < 1 \ . \label{KE}
\end{equation}

The eigenspeeds of the Whitham equations (\ref{KdVW}) are defined in terms of
$P_0$ and $P_1$ of (\ref{P01}), 
\begin{equation*}  
\lambda_i(u_1, u_2, u_3) = 12 \frac{
  P_1(u_i, u_1, u_2, u_3) }{P_0(u_i, u_1, u_2, u_3)} \ , \quad i=1,2,3 \ ,
\end{equation*}
which give 
\begin{align}
\lambda_1(u_1, u_2, u_3) &= 2 (u_1 + u_2 + u_3) + 4(u_1 - u_2)
\frac{K(s)}{E(s)} \ , \nonumber \\
\lambda_2(u_1, u_2, u_3) &= 2(u_1 + u_2 + u_3) + 4 (u_2 - u_1)
\frac{ sK(s) }{E(s) - (1-s) K(s)} \ , \label{lambda} \\
\lambda_3(u_1, u_2, u_3) &= 2(u_1 + u_2 + u_3) + 4(u_2-u_3)
\frac{K(s)}{E(s) - K(s)} \ . \nonumber 
\end{align}

Using (\ref{KE}), we obtain
\begin{eqnarray}
\lambda_{1} - 2(u_{1} + u_{2} + u_{3}) &>& 0 \ , \label{l1>}  \\
\lambda_{2} - 2(u_{1} + u_{2} + u_{3}) &<& 0 \ , \label{l2<}  \\
\lambda_{3} - 2(u_{1} + u_{2} + u_{3}) &<& 0 \ , \label{l3<}
\end{eqnarray}
for $u_1 > u_2 > u_3$. In view of (\ref{K}-\ref{E2}), we find that $\lambda_{1}$,
$\lambda_{2}$ and
$\lambda_{3}$ have behavior:

(1) At $u_{2}$ = $u_{3}$:
\begin{equation}
\label{tr}
\begin{array}{ll}
\lambda_{1}(u_1, u_2, u_3) = 6 u_{1} \ , \\
\lambda_{2}(u_1, u_2, u_3) =
\lambda_{3}(u_1, u_2, u_3) = 12 u_{3} - 6 u_{1} \ .
\end{array}
\end{equation}

(2) At $u_{1}$ = $u_{2}$:
\begin{equation}
\label{le}
\begin{array}{ll}
\lambda_{1}(u_1, u_2, u_3) =
\lambda_{2}(u_1, u_2, u_3) = 4 u_{1} + 2 u_{3} \ , \\
\lambda_{3}(u_1, u_2, u_3) = 6 u_{3} \ .
\end{array}
\end{equation}

The eigenspeeds of the Whitham equations (\ref{5KdVW})
are 
\begin{equation} \label{eq18a}
\mu_i(u_1, u_2, u_3) = 80 \frac{                                                                                                         
  P_2(u_i, u_1, u_2, u_3) }{P_0(u_i, u_1, u_2, u_3)} \ , \quad i=1,2,3 \ .                                                                                       
\end{equation}
They can be expressed in
terms of $\lambda_1$, $\lambda_2$ and $\lambda_3$ of
the KdV.

\pagebreak[2]

\begin{lem} \cite{Tian3} The eigenspeeds, $\mu_i$'s, satisfy: 
\begin{enumerate}[1.]
\item 
\begin{equation} \label{eq19}
\mu_i(u_1, u_2, u_3) = 
\frac{1}{2} \left[ \lambda_i - 2(u_1 + u_2 + u_3) \right]
\frac{\pd q}{\pd u_i}(u_1, u_2, u_3) + 
q(u_1, u_2, u_3) \ , 
\end{equation}
where $q(u_1, u_2, u_3)$ is the solution of the boundary value problem
of the Euler-Poisson-Darboux equations:
\begin{align} \label{eq20}
2 (u_i - u_j) \frac{\pd^2 q}{\pd u_i \pd u_j} &=
\frac{\pd q}{\pd u_i} - \frac{\pd q}{\pd u_j} \ , \quad i,j=1,2,3 \ ; i \neq j \ ,\\
q(u, u, u) &= 30 u^2 . \nonumber
\end{align}

\item
\begin{equation}
\label{a23}
{\pd \mu_i \over \pd u_j} = {{\pd \lambda_i \over \pd u_j} \over \lambda_i - \lambda_j} \ [\mu_i - \mu_j] \ ,       
\quad \quad i \neq j \ .
\end{equation}
\end{enumerate}

\end{lem}
The solution of (\ref{eq20}) is a symmetric quadratic function of $u_1$, $u_2$ and $u_3$
\begin{equation}
\label{q} q(u_1, u_2, u_3) = 6(u_1^2 + u_2^2 + u_3^2) + 4(u_1 u_2 + u_1 u_3 + u_2 u_3) \ .
\end{equation}

For KdV, $\lambda_i$'s satisfy \cite{Tian1} 
\begin{equation*}
\frac{\pd \lambda_3 }{\pd u_3} < \frac{3}{2}
\frac{\lambda_2 - \lambda_3 }{u_2 - u_3} < \frac{\pd
  \lambda_2}{\pd u_2}  
\end{equation*}
for $u_3 < u_2 < u_1 $. Similar results also hold for the fifth order KdV. 
\begin{lem}
\begin{eqnarray} 
\frac{\pd \mu_2}{\pd u_2} &>& \frac{3}{2}  \ \frac{\mu_2 - \mu_3}{u_2 - u_3}  \quad \mbox{if} \  \ 
\frac{\pd q}{\pd u_2} > 0 \label{ine1} \ , \\
\frac{\pd \mu_3}{\pd u_3} &<& \frac{3}{2}  \ \frac{\mu_2 - \mu_3}{u_2 - u_3}  \quad \mbox{if} \ \ 
\frac{\pd q}{\pd u_3} > 0 \label{ine2} \ ,
\end{eqnarray}
for $u_3 < u_2 < u_1$.
\end{lem}
\begin{proof}
We use (\ref{eq19}) to calculate 
\begin{align} \nonumber
\frac{\pd \mu_3}{\pd u_3} &= \frac{1}{2}
\frac{\pd \lambda_3}{\pd u_3} 
\frac{\pd q}{\pd u_3} + \frac{1}{2} [\lambda_3
-2(u_1 + u_2 + u_3)]\frac{\pd^2 q}{\pd u_3^2} \\
&< \frac{3}{4} \frac{\lambda_2 - \lambda_3}{u_2 - u_3}
\frac{\pd q}{\pd u_3} + \frac{1}{2} [\lambda_3 -
2(u_1 + u_2 + u_3)] \frac{\pd^2 q}{\pd u_3^2} \ , \label{eq20a} 
\end{align}
and 
\begin{align}
\mu_2 - \mu_3 &= \frac{1}{2} \left(
\lambda_2 - \lambda_3 \right) \frac{\pd q}{\pd
  u_3} + \frac{1}{2} [ \lambda_3 - 2(u_1 + u_2 + u_3)
] \left( \frac{\pd q}{\pd u_2} - \frac{\pd
  q}{\pd u_3}\right) \nonumber \\
&= \frac{1}{2} \left( \lambda_2 - \lambda_3 \right)
\frac{\pd q}{\pd u_3} + \frac{1}{2} [ \lambda_3 -
2(u_1 + u_2 + u_3)] 2 (u_2 - u_3) \frac{\pd^2 q}{\pd
  u_2 \pd u_3} \nonumber \\
&= \frac{2}{3} (u_2 - u_3) \left( \frac{3}{4} \frac{\lambda_2 -
  \lambda_3 }{u_2 - u_3} \frac{\pd q}{\pd u_3} 
+ \frac{3}{2} [\lambda_3 - 2 (u_1 + u_2 + u_3)]
\frac{\pd^2 q}{\pd u_2 \pd u_3} \right), \label{eq20b}
\end{align}
where we have used equation (\ref{eq20})   
\begin{displaymath} \label{eq21}  
\frac{\pd q}{\pd u_2} - \frac{\pd q}{\pd u_3} 
= 2(u_2 - u_3) \frac{\pd^2 q}{\pd u_2 \pd u_3} . 
\end{displaymath}
It follows from formula (\ref{q}) for $q$ that
\begin{equation*}
3 \frac{\pd^2 q}{\pd u_2 \pd u_3 } = \frac{\pd^2 q}{\pd u_3^2} \ , 
\end{equation*}
which, along with with (\ref{eq20a}) and (\ref{eq20b}), proves (\ref{ine1}).

Inequality (\ref{ine2}) can be proved in the same way.

\end{proof}

The following calculations are useful in the subsequent sections.

Using formula (\ref{eq19}) for $\mu_2$ and $\mu_3$ and formulae (\ref{lambda}) for
$\lambda_2$ and $\lambda_3$, we obtain
\begin{equation}
\label{M}
\mu_2(u_1, u_2, u_3) - \mu_3(u_1, u_2, u_3) = {2(u_2 - u_3) K \over (K-E)[E-(1-s)K]}
M(u_1, u_2, u_3) \ , 
\end{equation}
where
$$M(u_1, u_2, u_3) = [{\pd q \over \pd u_3} + (1-s) {\pd q \over \pd u_2}] E -(1-s)({\pd q \over \pd u_2} +
{\pd q \over \pd u_3}) K \ . $$
We then use (\ref{K3}), (\ref{E3}) and (\ref{q}) to calculate 
\begin{equation}
\label{pM}
{\pd M(u_1,u_2,u_3) \over \pd u_2} = {10(u_1 - 3u_2 - u_3) \over u_1 - u_3} \ (E - K) \ .
\end{equation}

We next consider
\begin{equation}
\label{F}
F(u_1, u_2,u_3) := {\mu_2(u_1, u_2, u_3)-\mu_3(u_1, u_2, u_3) \over u_2 - u_3} \ .
\end{equation}
Using formula (\ref{eq19}) for $\mu_2$ and $\mu_3$ and formulae (\ref{lambda}) for
$\lambda_2$ and $\lambda_3$, we obtain
\begin{eqnarray*}
F &=& - 2 {(1-s)K \over E - (1-s)K}
{\pd q \over \pd u_2} +
2 {K \over K - E} {\pd q \over \pd u_3} \\
&=& - 4 {s(1-s)K \over E - (1-s)K} (u_1 - u_3) {\pd^2 q \over \pd u_2 \pd u_3}
+ 2 [ {K \over K - E} - {(1-s)K \over E - (1-s)K}] {\pd q \over \pd u_3} \ ,
\end{eqnarray*}
where we have used equations (\ref{eq20}) in the last
equality. Finally, we use the expansions (\ref{K}-\ref{E}) for $K$ and $E$ to obtain
\begin{eqnarray}
F(u_1, u_2,u_3) &=& -4 [ (2 - {7 \over 4} s + \cdots ) (u_1 - u_3)
{\pd^2 q \over \pd u_2 \pd u_3} + (-{3 \over 4} + O(s^2)){\pd q \over \pd u_3}
] \nonumber \\
&=& - 16 [(2 - {7 \over 4} s + \cdots ) (u_1 - u_3)
+ (-{3 \over 4} + O(s^2)) (u_1 + u_2 + 3 u_3)] \ ,
\label{F2}
\end{eqnarray}
where we have used formula (\ref{q}) for $q$ in the last equality.

\section{A Self-similar Solution} 

In this section, we construct the self-similar solution of the Whitham equations 
(\ref{5KdVW}) for the initial function (\ref{step}) with $a=1$ and $b=0$. We will study all the other step
like initial data in Section 5.

\begin{thm}(see Figure 1.)
For the step-like initial data $u_0(x)$ of (\ref{step}) with $a=1, b=0$, the solution of the Whitham equations
(\ref{5KdVW}) is given by
\begin{equation}
\label{ws1}
u_1 = 1 \ , \quad x = \mu_2(1, u_2, u_3) \ t \ , \quad x = \mu_3(1, u_2, u_3) \ t  
\end{equation}
for $-15 t < x \leq \alpha t$ and by
\begin{equation}
\label{ws2}
u_1 = 1 \ , \quad x = \mu_2(1, u_2, 0) \ t \ , \quad u_3 = 0  
\end{equation}
for $\alpha t \leq x < 16 t$, where $\alpha = \mu_2(1, u^*, 0)$ and $u^*$ is the unique 
solution $u_2$ of $\mu_2(1,u_2,0)=\mu_3(1,u_2,0)$ in the interval $0<u_2<1$. 
Outside the region $-15 t < x < 16 t$, the solution of the Burgers
type equation (\ref{5Burgers}) is given by 
\begin{equation}
\label{bs1}  u \equiv 1 \quad \mbox{$x \leq - 15 t$}  
\end{equation} 
and
\begin{equation} 
\label{bs2}  u \equiv 0 \quad \mbox{$x \geq 16 t$} \ . 
\end{equation}
\end{thm}

The boundaries $x = - 15 t$ and $x = 16 t$ are called the trailing and leading edges, respectively.
They separate the solutions of the Whitham equations and Burgers type equations.
The Whitham solution matches the Burgers type solution in the following fashion (see Figure 1.):
\begin{eqnarray}
\label{tr1}
u_1 &=& \mbox{the Burgers type solution defined outside the region} \ , \\
\label{tr2}
u_2 &=& u_3 \ , 
\end{eqnarray}
at the trailing edge; 
\begin{eqnarray}
\label{le1}
u_1 &=& u_2 \ , \\
\label{le2}
u_3 &=& \mbox{the Burgers type solution defined outside the region} \ , 
\end{eqnarray}
at the leading edge. 

The proof of Theorem 3.1 is based on a series of lemmas.

We first show that the solutions defined by formulae (\ref{ws1}) and (\ref{ws2}) 
indeed satisfy the Whitham equations (\ref{5KdVW}) \cite{dub, tsa}. 

\begin{lem}

\begin{enumerate}
\item The functions $u_1$, $u_2$ and $u_3$ determined by equations (\ref{ws1})
give a solution of the Whitham equations (\ref{5KdVW}) as long as $u_2$ and $u_3$
can be solved from (\ref{ws1}) as functions of $x$ and $t$.

\item The functions $u_1$, $u_2$ and $u_3$ determined by equations (\ref{ws2})
give a solution of the Whitham equations (\ref{5KdVW}) as long as $u_2$ 
can be solved from (\ref{ws2}) as a function of $x$ and $t$.
\end{enumerate}

\end{lem}

\begin{proof}
 
(1) $u_1$ obviously satisfies the first equation of (\ref{5KdVW}). To verify the 
second and third equations, we observe that 
\begin{equation}
\label{dia}
\frac{\pd \mu_2 }{\pd u_3} = \frac{\pd \mu_3 }{\pd u_2} = 0
\end{equation}
on the solution of (\ref{ws1}). To see this, we use (\ref{a23}) to calculate
$$\frac{\pd \mu_2 }{\pd u_3} = {{\pd \lambda_2 \over \pd u_3} \over \lambda_2 - \lambda_3}
\ (\mu_2 - \mu_3) = 0 \ .$$
The second part of (\ref{dia}) can be shown in the same way.

We then calculate the partial derivatives of the second equation of (\ref{ws1})
with respect to $x$ and $t$.
$$ 1 = \frac{\pd \mu_2 }{\pd u_2} \ t u_{2x} \ , \quad 0 = \frac{\pd \mu_2 }{\pd u_2} \ t u_{2t} + \mu_2 \ ,$$ 
which give the second equation of (\ref{5KdVW}). 

The third equation of (\ref{5KdVW}) can be verified in the same way.

(2) The second part of Lemma 3.2 can easily be proved.

\end{proof}

We now determine the trailing edge. Eliminating $x$ and $t$ from the last two equations of (\ref{ws1})
yields 
\begin{equation}
\label{m23}
\mu_2(1, u_2, u_3) - \mu_3(1, u_2, u_3) = 0 \ . 
\end{equation}
Since it degenerates at $u_2 = u_3$, we replace (\ref{m23}) by 
\begin{equation}
\label{F1}
F(1, u_2,u_3) := {\mu_2(1, u_2, u_3)-\mu_3(1, u_2, u_3) \over u_2 - u_3} = 0 \ .
\end{equation}
Here, the function $F$ is also defined in (\ref{F}).

Therefore, at the trailing edge where $u_2=u_3$, i.e., $s=0$, equation
(\ref{F1}), in view of the expansion (\ref{F2}), becomes
$$2(1-u_3) - {3 \over 4} (1 + 4u_3) = 0 \ ,$$
which gives $u_2=u_3 = 1/4$.

\begin{lem}
Equation (\ref{F1}) has a unique solution satisfying $u_2=u_3$. The solution
is $u_2=u_3=1/4$. The rest of equations (\ref{ws1}) at the trailing edge
are $u_1=1$ and
$x/t = \mu_2(1, 1/4, 1/4) = -15$.
\end{lem}  

Having located the trailing edge, we now solve equations (\ref{ws1}) in the
neighborhood of the trailing edge. We first consider equation (\ref{F1}).
We use (\ref{F2}) to differentiate $F$ at the trailing edge $u_1=1$, $u_2=u_3=1/4$
$${\pd F(1, {1 \over 4}, {1 \over 4}) \over \pd u_2} = 
{\pd F(1, {1 \over 4}, {1 \over 4}) \over \pd u_3} = 40 \ ,$$
which show that equation (\ref{F1}) or equivalently (\ref{m23}) can be 
inverted to give $u_3$ as a decreasing
function of $u_2$
\begin{equation}
\label{A} u_3 = A(u_2)
\end{equation}
in a neighborhood of $u_2=u_3=1/4$. 

We now extend the solution (\ref{A}) of equation (\ref{m23}) in the region
$1 > u_2 > 1/4 > u_3 > 0$ as far as possible. We deduce from Lemma 2.2 that
\begin{equation}
\label{dia2}
{\pd \mu_2 \over \pd u_2} > 0 \ , \quad {\pd \mu_3 \over \pd u_3} <  0
\end{equation}
on the solution of (\ref{m23}).
Because of (\ref{dia}) and (\ref{dia2}), solution (\ref{A}) of equation (\ref{m23}) 
can be extended as long as $1 > u_2 > 1/4 > u_3 > 0$. 

There are two possibilities: (1) $u_2$ touches $1$ before or simultaneously
as $u_3$ reaches $0$ and (2) $u_3$ touches $0$ before $u_2$ reaches $1$.

It follows from (\ref{le}) and (\ref{eq19}) that
$$\mu_2(1,1,u_3) > \mu_3(1,1,u_3) \quad \mbox{for $0 \leq u_3 < 1$} \ .$$  
This shows that (1) is impossible. Hence, $u_3$ will touch $0$ before $u_2$
reaches $1$. When this happens, equation (\ref{m23}) becomes
\begin{equation}
\mu_2(1, u_2, 0) - \mu_3(1, u_2, 0) = 0 \ . \label{u2}
\end{equation}

\begin{lem}
Equation (\ref{u2}) has a simple zero in the region $0 < u_2 < 1$, counting
multiplicities. Denoting the zero by $u^*$, then $\mu_2(1, u_2, 0) - \mu_3(1, u_2, 0)$ is positive
for $u_2 > u^*$ and negative for $u_2 < u^*$.
\end{lem}

\begin{proof}
We now use (\ref{M}) and (\ref{pM}) to prove the lemma.
In equation (\ref{M}), $K-E$ and $E-(1-s)K$ are all positive for $0<s<1$ in view of (\ref{KE}).
By (\ref{pM}), 
$${\pd M(1, u_2, 0) \over \pd u_2} = 10(3u_2 -1)[K - E] \quad \ \mbox{for $0 < u_2 <1$} \ .$$
Since $M(1, u_2, 0)$ vanishes at $u_2=0$ and is positive at $u_2=1$ in view of (\ref{K}-\ref{E2}),
we conclude from the above derivative that $M(1, u_2, 0)$ has a simple zero in $0<u_2<1$. This zero is exactly
$u^*$ and the rest of the theorem can be proved easily.
\end{proof}

Having solved equation (\ref{m23}) for $u_3$ as a decreasing function of $u_2$
for $1/4 \leq u_2 \leq u^*$, we turn to equations (\ref{ws1}). Because of (\ref{dia2}),
the second equation of (\ref{ws1}) gives $u_2$ as a increasing function of $x/t$, for
$-15 \leq x/t \leq \alpha$, where 
\begin{equation}
\label{alpha} \alpha = \mu_2(1, u^*, 0).
\end{equation}
Consequently, $u_3$ is a decreasing function of $x/t$ in the same interval.

\begin{lem}
The last two equations of (\ref{ws1}) can be inverted to give $u_2$ and $u_3$ as 
increasing and decreasing functions, respectively, of the self-similarity variable
$x/t$ in the interval $-15 \leq x/t \leq \alpha$, where $\alpha = \mu_2(1, u^*, 0)$
and $u^*$ is given in Lemma 3.4.
\end{lem}


We now turn to equations (\ref{ws2}). We want to solve the second equation
when $x/t > \alpha$ or equivalently when $u_2 > u^*$. According to Lemma 3.4,
$\mu_2(1, u_2, 0) - \mu_3(1, u_2, 0) > 0$ for $u^* < u_2 < 1$, which, together
with (\ref{ine1}), shows that
$${\pd \mu_2(1, u_2, 0) \over \pd u_2} > 0 \ .$$
Hence, the second equation of (\ref{ws2}) can be solved for $u_2$ as an increasing
function of $x/t$ as long as $u^* < u_2 < 1$. When $u_2$ reaches $1$, we have
$$x/t = \mu_2(1, 1, 0) = 16 \ ,$$
where we 
have used (\ref{le}) and (\ref{eq19}) in the last equality. We have therefore proved the following result.

\begin{lem}
The second equation of (\ref{ws2}) can be inverted to give $u_2$ as an increasing
function of $x/t$ in the interval $\alpha \leq x/t \leq 16$.
\end{lem}

We are ready to conclude the proof of Theorem 3.1.
 
The Burgers type solutions (\ref{bs1}) and (\ref{bs2}) are trivial.

According to Lemma 3.5, the last two equations of (\ref{ws1}) determine $u_2$ 
and $u_3$ as functions of 
$x/t$ in the region $-15 \leq x/t \leq \alpha$. By the first part of Lemma 3.2, the
resulting $u_1$, $u_2$ and $u_3$ satisfy the Whitham equations (\ref{5KdVW}).
Furthermore, the boundary conditions (\ref{tr1}) and (\ref{tr2}) are satisfied
at the trailing edge $x = -15t$.

Similarly, by Lemma 3.6, the second equation of (\ref{ws2}) determines $u_2$ 
as a function of $x/t$ in the region $\alpha \leq x/t \leq 16$. It then follows from
the second part of Lemma 3.2 that $u_1$, $u_2$ and $u_3$ of (\ref{ws2}) satisfy 
the Whitham equations (\ref{5KdVW}).
They also satisfy the boundary conditions (\ref{le1}) and (\ref{le2}) at the
leading edge $x = 16t$. 

We have therefore completed the proof of Theorem 3.1.  

A graph of the
numerical solution of the Whitham equations is given in Figure
\ref{fig1}.

\section{The Minimization Problem}

The zero dispersion limit of the solution of the fifth order KdV 
equation (\ref{5KdV}) with step-like initial function (\ref{step}),
$a=1$, $b=0$, 
is also determined by
a minimization problem with constraints \cite{lax, lax2, ven}
\begin{equation}
\label{mini}
\underset{\{\psi \geq 0, \  \psi \in L^1 \}}
{\rm
Minimize}
\{ - \frac{1}{2 \pi} \int_0^1 \int_0^1 \log \Big|\frac{\eta - \mu}
{\eta + \mu
}\Big|
\psi(\eta) \psi(\mu) d \eta d \mu + \int_0^1 [\eta x - 16 \eta^5 t]
\psi(\eta) d \eta \} \ .
\end{equation}

In this section, we will use the self-similar solution of Section 3 to construct
the minimizer. We first define a linear operator
$$L \psi(\eta) = {1 \over 2 \pi} \int_0^1 log \left ( {\eta - \mu \over \eta + \mu} \right )^2
\psi(\mu) d \mu \ .$$
The variational conditions are
\begin{eqnarray}
L \psi = x \eta - 16 t \eta^5 \quad \mbox{where $\psi > 0$} \ ,  \label{con1} \\  
L \psi \leq x \eta - 16 t \eta^5 \quad \mbox{where $\psi = 0$}  \ . \label{con2}   
\end{eqnarray}
The constraint for the minimization problem is
\begin{equation}
\label{con}
\psi \geq 0 \ . 
\end{equation}

The minimizer of (\ref{mini}) is given explicitly:

\begin{thm}
The minimizer of the variational problem (\ref{mini}) is as follows:
\begin{enumerate}

\item For $x \leq -15 t$, 
\begin{equation*}
\psi(\eta) = {- x \eta + 80 t \eta (\eta^4 - {1 \over 2} \eta^2 - {1
    \over 8}) \over \sqrt{1 - \eta^2}} \ .
\end{equation*}

\item For $-15 t < x < \alpha t$, 
\begin{equation*}
\psi(\eta) = \left \{ \begin{matrix} - {-x \eta P_0(\eta^2, 1, u_2, u_3) + 80 t \eta P_2(\eta^2, 1, u_2, u_3)
\over \sqrt{(1 - \eta^2) (u_2 - \eta^2)(u_3 - \eta^2)}} & \quad \eta < \sqrt{u_3} \\
0 & \quad \sqrt{u_3} < \eta < \sqrt{u_2} \\ 
{-x \eta P_0(\eta^2, 1, u_2, 0) + 80 t \eta P_2(\eta^2, 1, u_2, 0) 
\over \sqrt{(1 - \eta^2) (\eta^2 - u_2)(\eta^2 - u_3)}} & \quad \sqrt{u_2} < \eta < 1 \end{matrix} \right. \ ,
\end{equation*}
where $P_0$ and $P_2$ are defined in (\ref{eq15}) and $u_2$ and $u_3$ are determined by equations (\ref{ws1}).

\item For $\alpha t < x < 16 t$,
\begin{equation*}
\psi(\eta) = \left \{ \begin{matrix} 0 & \eta < \sqrt{u_2} \\
{-x P_0(\eta^2, 1, u_2, 0) + 80 t P_2(\eta^2, 1, u_2, 0)
\over \sqrt{(1 - \eta^2) (\eta^2 - u_2)}} &  \sqrt{u_2} < \eta < 1 \end{matrix} \right. \ , 
\end{equation*}
where $u_2$ is determined by (\ref{ws2}).

\item For $x \geq 16t$, $$\psi(\eta) \equiv 0 \ .$$
\end{enumerate}

\end{thm}

\begin{proof}

We extend the function $\psi$ defined on $[0,1]$ to the entire real line by setting
$\psi(\eta) = 0$ for $\eta > 1$ and taking $\psi$ to be odd. In this way, the operator
$L$ is connected to the Hilbert transform $H$ on the real line \cite{lax}:
$$L \psi(\eta) = \int_0^{\eta} H \psi(\mu) d \mu \quad \mbox{where} \  H \psi (\eta) = {1 \over \pi}
P.V. \int_{- \infty}^{+ \infty} {\psi(\mu) \over \eta - \mu} d \mu \ .$$

We verify case (4) first. Clearly $\psi(\eta) = 0$ satisfies the constraints (\ref{con}).
We now check the variational conditions (\ref{con1}-\ref{con2}). Since $\psi=0$,
$$L \psi = 0 \leq x \eta - 16 t \eta^5 \ ,$$
where the inequality follows from $x \geq 16t$ and $0 \leq \eta \leq 1$. Hence,
variational conditions (\ref{con1}-\ref{con2}) are satisfied. 

Next we consider case (1). We write $\psi(\eta)$ as the real part of $g_1(\eta)$ for real $\eta$,
where 
$$g_1= \sqrt{-1}(x - 80 t \eta^4) + { \sqrt{-1} [-x \eta + 80 t \eta (\eta^4 - {1 \over 2} \eta^2 -
{1 \over 8})] \over \sqrt{\eta^2 -1}} \ .$$
The function $g_1$ is analytic in the upper
half complex plane $Im (\eta) > 0$ and $g_1(\eta) \approx O(1/\eta^2)$ for large $|\eta|$.
Hence, $H \psi (\eta) = Im [g_1(\eta)] = x - 80 t \eta^4$ on $0 \leq \eta \leq 1$, where $H$ is the
Hilbert transform \cite{lax}. We then have for $0 \leq \eta \leq 1$
$$L \psi (\eta) = \int_0^{\eta} H \eta(\mu) d \mu = x \eta - 16t \eta^5 \ ,$$
which shows that the variational conditions are satisfied. Since
$15 + 80 (\eta^4 - {1 \over 2} \eta^2 - {1 \over 8}) = 80 (\eta^2 - {1 \over 4})^2 \geq 0$,
it follows from $x \leq -15 t$ that $\psi \geq 0$. 
Hence, the constraint (\ref{con}) is verified.

We now turn to case (2). By Lemma 3.5, the last two equations of (\ref{ws1}) determine
$u_2$ and $u_3$ as functions of the self-similarity variable $x/t$ in the interval
$-15 \leq x/t \leq \alpha$. 

We write $\psi = Re \left(g_2\right) $ for real $\eta$, where
$$g_2 = \sqrt{-1}(x - 80 t \eta^4) + { \sqrt{-1} [-x \eta P_0(\eta^2, 1, u_2, u_3) + 80 t \eta P_2(\eta^2, 1, u_2, u_3)
]  \over \sqrt{\eta^2 -1)(\eta^2 - u_2)(\eta^2 - u_3)}} \ .$$
The function $g_2$ is analytic in $Im (\eta) > 0$ and $g_2(\eta) \approx O(1/\eta^2)$ for large $|\eta|$
in view of the asymptotics (\ref{eq16}) for $P_0$ and $P_2$. Hence, taking the imaginary part of $g_2$
yields
$$ H \psi(\eta) = \left \{ \begin{matrix} x - 80 t \eta^4 & 0 < \eta < \sqrt{u_3} \\
       x - 80 t \eta^4 - {[-x P_0(\eta^2, 1, u_2, 0) + 80 t P_2(\eta^2, 1, u_2, 0)] \eta
\over \sqrt{(1 - \eta^2) (u_2 - \eta^2)(\eta^2 - u_3)}} &  \sqrt{u_3} < \eta < \sqrt{u_2} \\
 x - 80 t \eta^4 & \sqrt{u_2} < \eta < 1 \end{matrix} \right. \ .$$ 
We then have 
$$L \psi (\eta) =  \left \{ \begin{matrix} x \eta - 16 t \eta^5 & 0 < \eta < \sqrt{u_3} \\
 x \eta - 16 t \eta^5 - \int_{\sqrt{u_3}}^{\eta} {[-x P_0 + 80 t P_2] \mu \over \sqrt{(1 - \mu^2)
(u_2 - \mu^2) (\mu^2 - u_3)} } d \mu & \sqrt{u_3} < \eta < \sqrt{u_2} \\
 x \eta - 16 t \eta^5 & \sqrt{u_2} < \eta < 1  \end{matrix} \right. \ ,$$
where we have used 
\begin{equation}
\label{loop}
\int_{\sqrt{u_3}}^{\sqrt{u_2}} {[-x P_0 + 80 t P_2] \mu \over \sqrt{(1 - \mu^2)
(u_2 - \mu^2) (\mu^2 - u_3)} } d \mu = 0 \ ,
\end{equation}
which is a consequence of (\ref{eq17}) for $P_0$ and $P_2$.  

We study the zeros of $-x P_0 + 80 t P_2$. It has two zeros
at $\eta = \sqrt{u_2}$ and $\eta = \sqrt{u_3}$. This follows from (\ref{eq18a}) and (\ref{ws1}).   
It also has a zero between $\sqrt{u_2}$ and $\sqrt{u_3}$ because of (\ref{loop}).
Since it is a cubic polynomial of $\eta^2$, $-x P_0 + 80 t P_2$ has no more than three
zeros on the positive $\eta$ axis and furthermore these three positive zeros are simple.

Since the leading term in $-x P_0 + 80 t P_2$ is $80t \eta^6$, the polynomial is positive for 
$\eta > \sqrt{u_2}$ and negative for $0 \leq \eta < \sqrt{u_3}$. This proves $\psi \geq 0$; so
(\ref{con}) is verified. 
Since $-x P_0 + 80 t P_2$ changes sign at each simple zero, it follows from (\ref{loop}) that
$$\int_{\sqrt{u_3}}^{\eta} {[-x P_0 + 80 t P_2] \mu \over \sqrt{(1 - \mu^2)
(u_2 - \mu^2) (\mu^2 - u_3)} } d \mu > 0$$
for $\sqrt{u_3} < \eta < \sqrt{u_2}$.    
This verifies the variational conditions (\ref{con1}) and (\ref{con2}).

We finally consider case (3). By Lemma 3.6, the second equation of (\ref{ws2}) determines $u_2$
as an increasing function of $x/t$ in the interval $\alpha \leq x/t \leq 16$.

We write $\psi = Re (g_3)$ for real $\eta$, where
$$g_3 = \sqrt{-1}(x - 80 t \eta^4) + { \sqrt{-1} [-x P_0(\eta^2, 1, u_2, 0) + 80 t P_2(\eta^2, 1, u_2, 0)
]  \over \sqrt{(\eta^2 -1)(\eta^2 - u_2)}} \ .$$
The function $g_3$ is analytic in $Im (\eta) > 0$ and $g_3(\eta) \approx O(1/\eta^2)$ for large $|\eta|$
in view of the asymptotics (\ref{eq16}) for $P_0$ and $P_2$. Hence, taking the imaginary part of $g_3$
yields
$$H \psi(\eta) = \left \{ \begin{matrix} x - 80 t \eta^4 - {-x P_0(\eta^2, 1, u_2, 0) + 80 t P_2(\eta^2, 1, u_2, 0)
\over \sqrt{(1 - \eta^2)(u_2 - \eta^2)}} & 0 < \eta < \sqrt{u_2} \\
x - 80 t \eta^4  & \sqrt{u_2} < \eta < 1 \end{matrix} \right. \ .$$
We then have
$$L \psi (\eta) = \left \{ \begin{matrix} x \eta - 16t \eta^5 - \int_0^{\eta} {-x P_0 + 80 t P_2 
\over \sqrt{(1 - \mu^2)(u_2 - \mu^2)}} d \mu & 0< \eta < \sqrt{u_2} \\
 x \eta - 16t \eta^5 & \sqrt{u_2} < \eta < 1 \end{matrix} \right. \ ,$$
where we have used
\begin{equation}
\label{loop2} \int_0^{\sqrt{u_2}} {-x P_0(\mu^2, 1, u_2, 0) + 80 t P_2(\mu^2, 1, u_2, 0) \over 
\sqrt{(1 - \mu^2)(u_2 - \mu^2)}} d \mu = 0 \ , 
\end{equation}
which is a consequence of (\ref{eq17}) for $P_0$ and $P_2$.  
 
The function $-x P_0(\eta^2, 1, u_2, 0) + 80 t P_2(\eta^2, 1, u_2, 0)$ has two zeros on the positive
$\eta$-axis. One is at
$\eta = \sqrt{u_2}$, in view of (\ref{eq18a}) and (\ref{ws2}). The other is between $0$ and $\sqrt{u_2}$,
in view of (\ref{loop2}). At $\eta=0$, the function has a positive value. To see this, 
\begin{equation}
\label{P00}
-x P_0(0, 1, u_2, 0) + 80 t P_2(0, 1, u_2, 0) = P_0(0, 1, u_2, 0)[-x + t \mu_3(1, u_2, 0)] \ . \\
\end{equation}
According to Lemma 3.4, $\mu_2(1, u_2, 0) > \mu_3(1, u_2, 0)$ when $u_2 > u^*$ or equivalently when 
$\alpha < x/t < 16$. It follows from formula (\ref{P01}) and inequality (\ref{KE}) that $P_0(0,1,u_2,0) < 0$.
Hence, the right hand side of (\ref{P00}) is bigger than
$$P_0(0, 1, u_2, 0)[-x + t \mu_2(1, u_2, 0)] = 0 \ ,$$
where the equality comes from (\ref{ws2}). Since it is a cubic polynomial in $\eta^2$ and 
since it is positive for large $\eta > 0$, 
the function $-x P_0(\eta^2, 1, u_2, 0) + 80 t P_2(\eta^2, 1, u_2, 0)$ can have at most two zeros
on the positive $\eta$-axis. Hence, the above two zeros are all simple zeros. 

It now becomes straight forward to check the variational conditions (\ref{con1}-\ref{con2})
and the constraint (\ref{con}), just as we do in case (2).

\end{proof} 

\section{Other Step Like Initial Data}

In this section, we will classify all types of step like initial data (\ref{step}) for equation (\ref{5KdV}). 
When $a=0$, since 
$b \neq 0$, the solution of (\ref{5Burgers}) will never develop a shock. We therefore study the
cases $a > 0$ and $a < 0$. 
In the former case,
it is easy to check that, when $b> a$, the solution of equation
(\ref{5Burgers}) will never develop a shock; accordingly, we will restrict to $b < a$. Similarly,
in the latter case, we will confine ourselves to $b > a$.

We will only present our proofs briefly, since they are, more or less, similar to those in Section 3.

\subsection{Type I: $a > 0$, $a/4 \leq b < a$}

\begin{figure}[h] \label{fig2}
\begin{center}
\resizebox{12cm}{5cm}{\includegraphics{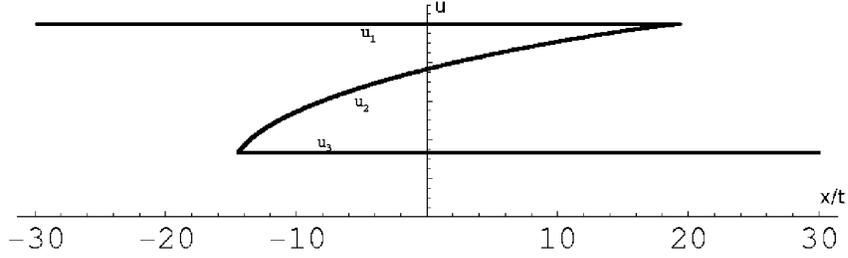}}
\caption{Self-Similar solution of the Whitham equations for
  $a=1$ and $b= 1/3$ of type I.}
\end{center}\end{figure}

\begin{thm}(see Figure 2.)
For the step like initial data (\ref{step}) with $a > 0$, $a/4 \leq b < a$, the solution of the
Whitham equations (\ref{5KdVW}) is given by 
\begin{equation*}
u_1 =a \ , \quad x = \mu_2(a, u_2, b) \ t \ , \quad u_3 = b
\end{equation*}
for $\mu_2(a,b,b) < x/t < \mu_2(a,a,b)$, where $\mu_2(a,b,b) = -10 a^2 - 40 ab + 80 b^2$ and
$\mu_2(a,a,b) = 16 a^2 + 8ab + 6 b^2$. Outside this interval, the solution of
(\ref{5Burgers}) is given by 
\begin{equation*}
u \equiv a \quad \mbox{$x/t \leq \mu_2(a,b,b)$}
\end{equation*}
and
\begin{equation*}
u \equiv b \quad \mbox{$x/t \geq \mu_2(a,a,b)$} \ .
\end{equation*}
\end{thm}

\begin{proof}

It suffices to show that $\mu_2(a, u_2, b)$ is an increasing function of $u_2$ for 
$b < u_2 < a$. By (\ref{pM}),
we have
$${d M(a, u_2, b) \over d u_2} = {10(3 u_2 + b -a) \over a - b} [K - E] > 0 $$
for $b<u_2<a$, where we have used $a/4 \leq b < a$ in the inequality. Since $M(a,u_2,b) = 0$
at $u_2 = b$, this implies that
$M(a,u_2,b) > 0$
for $b < u_2 < a$. It then follows from (\ref{M}) that $\mu_2(a, u_2, b) - \mu_3(a, u_2, b) > 0$.
By Lemma 2.2, we conclude that 
$${d \mu_2(a, u_2, b) \over d u_2} > 0$$ for $b < u_2 < a$.

\end{proof}

\subsection{Type II: $a > 0$, $-2a/3 < b < a/4$} 

\hfill

Theorem 3.1 is a special case of the following theorem.

\begin{thm} (see Figure 1.)
For the step like initial data (\ref{step}) with $a > 0$, $-2a/3 < b < a/4$, 
the solution of the Whitham equations
(\ref{5KdVW}) is given by
\begin{equation*}
u_1 = a \ , \quad x = \mu_2(a, u_2, u_3) \ t \ , \quad x = \mu_3(a, u_2, u_3) \ t
\end{equation*}
for $-15 a^2 < x/t \leq \mu_2(a, u^{**}, b)$ and by
\begin{equation*}
u_1 = a \ , \quad x = \mu_2(a, u_2, b) \ t \ , \quad u_3 = b
\end{equation*}
for $\mu_2(a, u^{**}, b) \leq x/t < 16a^2 + 8ab + 6 b^2$, where $u^{**}$ is the unique solution
$u_2$ of $\mu_2(a,u_2,b)=\mu_3(a,u_2,b)$ in the interval $b<u_2<a$.
Outside the region $-15 a^2 < x/t < 16 a^2 + 8ab + 6 b^2$, the solution of the Burgers
type equation (\ref{5Burgers}) is given by
\begin{equation*}
u \equiv a \ \quad \mbox{$x/t \leq -15 a^2$} 
\end{equation*}
and
\begin{equation*}
u \equiv b \ \quad \mbox{$x/t \geq 16a^2 + 8ab + 6 b^2$} \ .
\end{equation*}
\end{thm}

\begin{proof}

The trailing edge is determined by 
\begin{equation}
\label{F3}
F(a, u_2, u_3) = 0
\end{equation}
when $u_2 = u_3$. Here $F$ is given by (\ref{F}). In view of the expansion (\ref{F2}), the above equation
when $u_2=u_3$, i.e., $s=0$, reduces to 
$$2(a-u_3) - {3 \over 4}(a + 4u_3) = 0 \ ,$$
which gives $u_2=u_3=a/4$ at the trailing edge.

Having located the trailing edge, we solve equation (\ref{F3}) in the neighborhood of $u_2=u_3=a/4$. 
We use the expansion (\ref{F2}) to calculate
$${\pd F(a, {a \over 4}, {a \over 4}) \over \pd u_2} = {\pd F(a, {a \over 4}, {a \over 4}) \over \pd u_3} = 40 \ ,$$
which implies that equation (\ref{F3}) can be solved for $u_3$ as a decreasing function of $u_2$
near $u_2=u_3=a/4$. 

The solution of 
\begin{equation}
\label{ab}
\mu_2(a,u_2,u_3) - \mu_3(a,u_2,u_3)=0 
\end{equation}
can be extended as long as
$a > u_2 > a/4 > u_3 > b$. To see this, we need to show that
$${\pd \mu_2(a,u_2,u_3) \over \pd u_3} = 0 \ , \quad {\pd \mu_3(a,u_2,u_3) \over \pd u_2} = 0 \ ,
\quad {\pd \mu_2(a,u_2,u_3) \over \pd u_2} > 0 \ , \quad {\pd \mu_3(a,u_2,u_3) \over \pd u_3} < 0$$
on the solution of (\ref{ab}). The proof of the equalities is the same as that of (\ref{dia}) in Section 3.
To prove the inequalities, in view of Lemma 2.2, it is enough to
show that 
$${\pd q(a,u_2,u_3) \over \pd u_2} > 0 \ , \quad {\pd q(a,u_2,u_3) \over \pd u_3} > 0 \ .$$ 
We use formulae (\ref{eq19}) to rewrite equation (\ref{ab}) as
$${1 \over 2} [\lambda_2 - 2(a + u_2 + u_3)] {\pd q(a,u_2,u_3) \over \pd u_2}
= {1 \over 2} [\lambda_3 - 2(a + u_2 + u_3)] {\pd q(a,u_2,u_3) \over \pd u_3} \ ,$$
which, together with inequalities (\ref{l2<}) and (\ref{l3<}), proves that
${\pd q \over \pd u_2}$ and ${\pd q \over \pd u_3}$ have the same sign on the solution of
(\ref{ab}). On the other hand, we calculate from (\ref{q})
$${\pd q(a,u_2,u_3) \over \pd u_2} = 4 (a + 3 u_2 + u_3) > 0$$
for $a>u_2 > a/4 > u_3 > b > -2a/3$.  

We now extend the solution of (\ref{ab}) as far as possible in the region $a > u_2 > a/4 > u_3 > b$.
There are two possibilities: (1) $u_2$ touches $a$ before or simultaneously
as $u_3$ reaches $b$ and (2) $u_3$ touches $b$ before $u_2$ reaches $a$.

Possibility (1) is impossible. To see this, we
use (\ref{le}) and (\ref{eq19}) to calculate
\begin{equation}
\label{mu23}
\mu_2(a,a,u_3) - \mu_3(a,a,u_3) = 2(a - u_3) {\pd q(a,a,u_3) \over \pd u_3} = 8(a-u_3)(2a + 3u_3) \ , 
\end{equation}
which, in view of $b > -2a/3$, is positive for $b \leq u_3 < a$. 

Therefore, $u_3$ will touch $b$ before $u_2$ reaches $a$. When this happens, we have 
$\mu_2(a,u_2,b) - \mu_3(a,u_2,b) =0$. In the same way as we prove Lemma 3.4, we can show that
this equation has a unique solution
$u_2$ in the interval $b < u_2 < a$.

The rest of the proof is similar to that of Theorem 3.1.

\end{proof}

\subsection{Type III: $a > 0$, $b = -2a/3$}

\begin{figure}[h] \label{fig3}
\begin{center}
\resizebox{12cm}{5cm}{\includegraphics{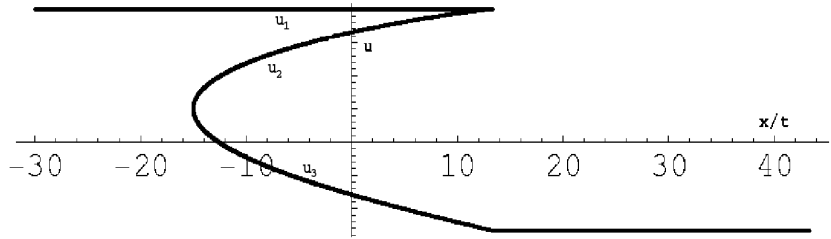}}
\caption{Self-Similar solution of the Whitham equations for
  $a=1$ and $b=- 2/3$ of type III.}
\end{center}\end{figure}

\begin{thm} (see Figure 3.)
For the step like initial data (\ref{step}) with $a > 0$, $b = -2a/3$,
the solution of the Whitham equations
(\ref{5KdVW}) is given by
\begin{equation*}
u_1 = a \ , \quad x = \mu_2(a, u_2, u_3) \ t \ , \quad x = \mu_3(a, u_2, u_3) \ t
\end{equation*}
for $-15 a^2 < x/t < 40 a^2/3$.
Outside the region, the solution of the Burgers
type equation (\ref{5Burgers}) is given by
\begin{equation*}
u \equiv a \quad \mbox{$x/t \leq -15 a^2$} 
\end{equation*}
and
\begin{equation*}
u \equiv b \quad \mbox{$x/t \geq 40 a^2/3$} \ .
\end{equation*}
\end{thm}

\begin{proof}

It suffices to show that $u_2$ and $u_3$ of $\mu_2(a, u_2, u_3) - \mu_3(a, u_2, u_3)=0$
reaches $a$ and $b=-2a/3$, respectively, simultaneously. To see this, we deduce from
equation (\ref{mu23}) that 
\begin{equation}
\label{mu23'} \mu_2(a,a,-2a/3) - \mu_3(a,a,-2a/3) = 8(a-2a/3)[2a + 3(-2a/3)] =0 \ .
\end{equation}
\end{proof}

\subsection{Type IV: $a > 0$, $b < -2a/3$}

\begin{figure}[h] \label{fig4}
\begin{center}
\resizebox{12cm}{4cm}{\includegraphics{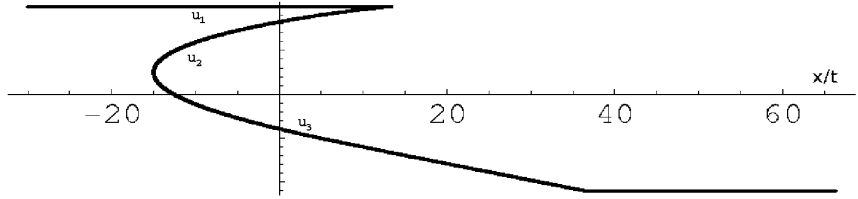}}
\caption{Self-Similar solution of the Whitham equations for
  $a=1$ and $b=-1.1$ of type IV.}
\end{center}\end{figure}

\begin{thm}(see Figure 4.)
For the step like initial data (\ref{step}) with $a > 0$, $b < -2a/3$,
the solution of the Whitham equations
(\ref{5KdVW}) is given by
\begin{equation*}
u_1 = a \ , \quad x = \mu_2(a, u_2, u_3) \ t \ , \quad x = \mu_3(a, u_2, u_3) \ t
\end{equation*}
for $-15 a^2 < x/t < 40 a^2 /3$.
Outside the region, the solution of the Burgers
type equation (\ref{5Burgers}) is given by
\begin{equation*}
u \equiv a \quad \mbox{$x/t \leq -15 a^2$} 
\end{equation*}
and
\begin{equation*}
u = \left\{ \begin{matrix} - \sqrt{{x \over 30 t}} 
& 40 a^2/3 \leq x/t \leq 30b^2 \\
b & x/t \geq 30b^2 \end{matrix} \right. \ . 
\end{equation*}
\end{thm}

\begin{proof}

By the calculation (\ref{mu23'}), when $u_2$ of $\mu_2(a, u_2, u_3) - \mu_3(a, u_2, u_3)=0$
touches $a$, the corresponding $u_3$ reaches $-2a/3$, which is above $b$. Hence, equations 
$$x = \mu_2(a, u_2, u_3) \ t \ , \quad x = \mu_3(a, u_2, u_3) \ t$$
can be inverted to give $u_2$ and $u_3$ as functions of $x/t$ in the region 
$\mu_2(a, a/4, a/4) < x/t \leq \mu_2(a, a, -2a/3)$.  To the right of this region,
the Burgers type equation (\ref{5Burgers}) has a rarefaction wave solution.

\end{proof}

\subsection{Type V: $a < 0$, $ b \leq - a/4 $}

\begin{figure}[h] \label{fig5}
\begin{center}
\resizebox{12cm}{5cm}{\includegraphics{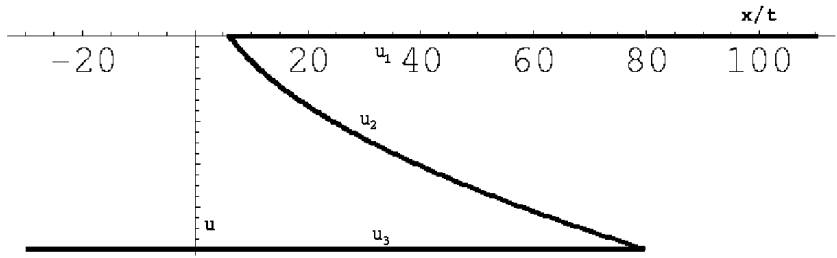}}
\caption{Self-Similar solution of the Whitham equations for
  $a=-1$ and $b= 0$ of type V.}
\end{center}\end{figure}

\begin{thm}(see Figure 5.)
For the step like initial data (\ref{step}) with $a < 0$, $a < b \leq - a/4$, the solution of the
Whitham equations (\ref{5KdVW}) is given by
\begin{equation*}
u_1 =b \ , \quad x = \mu_2(b, u_2, a) \ t \ , \quad u_3 = a
\end{equation*}
for $\mu_2(b,b,a) < x/t < \mu_2(b,a,a)$, where $\mu_2(b,b,a) = 6a^2 + 8ab + 16b^2$ and
$\mu_2(b,a,a) = 80a^2 - 40ab - 10 b^2$. Outside this interval, the solution of
(\ref{5Burgers}) is given by
\begin{equation*}
u \equiv a \quad \mbox{$x/t \leq \mu_2(b,b,a)$} 
\end{equation*}
and
\begin{equation*}
u \equiv b \quad \mbox{$x/t \geq \mu_2(b,a,a)$} \ .
\end{equation*}
\end{thm}

\begin{proof}

It suffices to show that $\mu_2(a, u_2, b)$ is a decreasing function of $u_2$ for
$a < u_2 < b$. By (\ref{eq19}),
we have
$${\pd \mu_2(b, u_2, a) \over \pd u_2} = {1 \over 2} {\pd \lambda_2 \over \pd u_2} 
{\pd q \over \pd u_2} + {1 \over 2} [\lambda_2 - 2(b + u_2 + a)] {\pd^2 q \over \pd u_2^2} \ .$$
The second term is negative because of (\ref{l2<}) and ${\pd^2 q \over \pd u_2^2} = 12 > 0$.
The first term is also negative. Its first factor is positive in view of (\ref{eq11}). 
The second factor
$${\pd q \over \pd u_2} = 4(b + 3 u_2 +a) < 0$$
for $a < u_2 < b$ because of $b \leq - a/4$.

\end{proof}

\subsection{Type VI: $a <0 $, $-a/4 < b < -2a$}

\begin{figure}[h] \label{fig6}
\begin{center}
\resizebox{12cm}{5cm}{\includegraphics{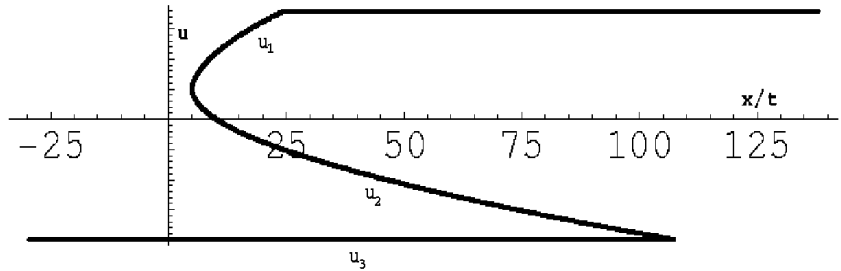}}
\caption{Self-Similar solution of the Whitham equations for
  $a= -1$ and $b= 1.2$ of type VI.}
\end{center}\end{figure}

\begin{thm}(see Figure 6.)
For the step like initial data (\ref{step}) with $a <0 $, $-a/4 < b < -2a$,
the solution of the Whitham equations
(\ref{5KdVW}) is given by
\begin{equation}
\label{ws1''}
x = \mu_1(u_1, u_2, a) \ t \ , \quad x = \mu_2(u_1, u_2, a) \ t \ , \quad u_3 = a
\end{equation}
for $5 a^2 < x/t \leq \mu_2(b, u^{***}, a)$ and by
\begin{equation}
\label{ws2''}
u_1 = b \ , \quad x = \mu_2(b, u_2, a) \ t \ , \quad u_3 = a
\end{equation}
for $\mu_2(b, u^{***}, a) \leq x/t < 80a^2 - 40ab - 10 b^2$, where $u^{***}$ is the unique solution
$u_2$ of $\mu_1(b,u_2,a)=\mu_2(b,u_2,a)$ in the interval $a<u_2<b$.
Outside the region $ 5 a^2 < x/t < 80a^2 - 40ab - 10 b^2$, the solution of the Burgers
type equation (\ref{5Burgers}) is given by
\begin{equation*}
u \equiv a \ \quad \mbox{$x/t \leq 5 a^2$} 
\end{equation*}
and
\begin{equation*}
u \equiv b \ \quad \mbox{$x/t \geq 80a^2 - 40ab - 10 b^2$} \ .
\end{equation*}
\end{thm}

\begin{proof}

We first locate the ``leading'' edge, i.e., the solution of equation (\ref{ws1''}) at $u_1=u_2$.
Eliminating $x/t$ from the first two equations of (\ref{ws1''}) yields
\begin{equation}
\label{mu12}
\mu_1(u_1,u_2,a) - \mu_2(u_1,u_2,a) = 0 \ .
\end{equation}
Since it degenerates at $u_1=u_2$, we replace (\ref{mu12}) by
\begin{equation}
\label{G}
G(u_1,u_2,a) := {\mu_1(u_1,u_2,a) - \mu_2(u_1,u_2,a) \over (u_1 - u_2) K(s)} = 0 \ .
\end{equation}
Using formulae (\ref{eq19}) for $\mu_1$ and $\mu_2$ and formulae (\ref{lambda}) for
$\lambda_1$ and $\lambda_2$, we write
$$G(u_1,u_2,a) =  {2 \over E [E - (1-s)K] } \{ ({\pd q \over \pd u_1} + s {\pd q \over \pd u_2}) E
- (1-s) {\pd q \over \pd u_1} K \} \ .$$
In view of (\ref{K2}) and (\ref{E2}), equation (\ref{G}) reduces to
$${\pd q(u_1,u_2,a) \over \pd u_1} + {\pd q(u_1,u_2,a) \over \pd u_2} = 0 $$
at the ``leading'' edge $u_1=u_2$. This gives
$$u_1 = u_2 = - {a \over 4} \ .$$

Having located the ``leading'' edge, we solve equation (\ref{G}) near $u_1 = u_2 = - a/4$.
We calculate 
$${\pd G(-a/4, -a/4, a) \over \pd u_1} = {\pd G(-a/4, -a/4, a) \over \pd u_2} = 32 \ .$$
These show that equation (\ref{G}) gives $u_1$ as a decreasing function of $u_2$ 
\begin{equation}
\label{B}
u_1 = B(u_2)
\end{equation}
in a neighborhood of $u_1 = u_2 = - a/4$.

We now extend the solution (\ref{B}) of equation (\ref{mu12}) as far as possible in the region 
$a < u_2 < -a/4 < u_1 < b$. We use formula (\ref{eq19}) to calculate
\begin{eqnarray*}
{\pd \mu_1 \over \pd u_1} &=& {1 \over 2} {\pd \lambda_1 \over \pd u_1}{\pd q \over \pd u_1}   
+ {1 \over 2} [\lambda_1 - 2(u_1 + u_2 + a)] {\pd^2 q \over \pd u_1^2} \ , \\
{\pd \mu_2 \over \pd u_2} &=& {1 \over 2} {\pd \lambda_2 \over \pd u_2}{\pd q \over \pd u_2}  
+ {1 \over 2} [\lambda_2 - 2(u_1 + u_2 + a)] {\pd^2 q \over \pd u_2^2} \ .
\end{eqnarray*}
In view of (\ref{eq11}), (\ref{l1>}) and (\ref{l2<}), we have
\begin{eqnarray*}
{\pd \mu_1 \over \pd u_1} &>& 0 \quad \mbox{if} \ \  {\pd q \over \pd u_1} > 0 \ , \\
{\pd \mu_2 \over \pd u_2} &<& 0 \quad \mbox{if} \ \ {\pd q \over \pd u_2} < 0 \ .
\end{eqnarray*}

We claim that 
\begin{equation}
\label{cl}
{\pd q \over \pd u_1} > 0 \ , \quad {\pd q \over \pd u_2} < 0
\end{equation}
on the 
solution of (\ref{mu12}) in the region $a < u_2 < -a/4 < u_1 < b$. To see this,
we use formula (\ref{eq19}) to rewrite equation (\ref{mu12}) as
$${1 \over 2} [\lambda_1 - 2(u_1 + u_2 + a)] {\pd q \over \pd u_1}
= {1 \over 2} [\lambda_2 - 2(u_1 + u_2 + a)] {\pd q \over \pd u_2} \ .$$
This, together with
$${\pd q \over \pd u_1} - {\pd q \over \pd u_2} = 2(u_1 - u_2) {\pd^2 q \over \pd u_1 \pd u_2}
= 8(u_1 - u_2) > 0 $$
for $u_1 > u_2$, and inequalities (\ref{l1>}) and (\ref{l2<}),
proves (\ref{cl}).

Hence, the solution (\ref{B}) can be extended as long as $a < u_2 < -a/4 < u_1 < b$.

There are two possibilities: (1) $u_1$ touches $b$ before 
$u_2$ reaches $a$ and (2) $u_2$ touches $a$ before or simultaneously as $u_1$ reaches $a$.

Possibility (2) is impossible. To see this, we
use (\ref{tr}), (\ref{eq19}) and (\ref{q}) to calculate
\begin{equation}
\label{2}
\mu_1(u_1,a,a) - \mu_2(u_1,a,a) = 40(u_1 -a)(u_1 + 2a) \ ,
\end{equation}
which is negative for $-a/4 < u_1 \leq b < -2a$.

Therefore, $u_1$ will touch $b$ before $u_2$ reaches $a$. When this happens, we have
\begin{equation}
\label{mu12e}
\mu_1(b,u_2,a) - \mu_2(b,u_2,a) =0 \ .
\end{equation}

\begin{lem}
Equation (\ref{mu12e}) has a simple zero, counting multiplicities, in the interval
$a < u_2 < b$. Denoting this zero by $u^{***}$, then $\mu_1(b,u_2,a) - \mu_2(b,u_2,a)$
is positive for $u_2 > u^{***}$ and negative for $u_2 < u^{***}$.
\end{lem}

\begin{proof}

We write
\begin{equation}
\label{mu12e'}
\mu_1(b,u_2,a) - \mu_2(b,u_2,a) = {2 (b - u_2) K \over E [E - (1-s)K] } \{ ({\pd q \over \pd u_1} 
+ s {\pd q \over \pd u_2} ) E - (1-s) {\pd q \over \pd u_1} K \} \ .
\end{equation}
Denote the parenthesis of (\ref{mu12e'}) by $N(b,u_2,a)$. Since $E - (1-s)K > 0$ for $a < u_2 < b$, 
the left hand side has a zero iff $N(b,u_2,a)$  
on the right has one. 

We now calculate 
\begin{equation*} 
{\pd N(b,u_2,a) \over \pd u_2} = {30 E(s) \over b - a} [u_2 - {a - b
\over 3}] \ .
\end{equation*}
Since $N(b,u_2,a)$ is zero at $u_2 = a$ and positive at
$u_2 = b$, we conclude from the above derivative that $N(b,u_2,a)$ has a simple zero
in $a < u_2 < b$.
\end{proof}

We now continue to prove Theorem 5.6. Having solved equation (\ref{mu12}) for $u_1$ as a
decreasing function of $u_2$ for $u^{***} < u_2 < - a/4$, we can then use the last two equations
of (\ref{ws1''}) to determine $u_1$ and $u_2$ as functions of $x/t$ in the interval
$\mu_2(-a/4, -a/4, a) < x/t < \mu_2(b, u^{***}, a)$.

We finally turn to equations (\ref{ws2''}). We want to solve the second equation
of (\ref{ws2''}), $x/t = \mu_2(b, u_2, a)$, for $u_2 < u^{***}$. It is enough to show that
$\mu_2(b, u_2, a)$ is a decreasing function of $u_2$ for $u_2 < u^{***}$.

According to Lemma 5.7, $\mu_1(b,u_2,a) - \mu_2(b,u_2,a) <  0$ for $u_2 < u^{***}$.
Using formula (\ref{eq19}) for $\mu_1$ and $\mu_2$, we have
$${1 \over 2} [\lambda_1 - 2(b + u_2 + a)] {\pd q \over \pd u_1}
< {1 \over 2} [\lambda_2 - 2(b + u_2 + a)] {\pd q \over \pd u_2} \ .$$
This, together with
$${\pd q \over \pd u_1} - {\pd q \over \pd u_2} = 2(b - u_2) {\pd^2 q \over \pd u_1 \pd u_2}
= 8(b - u_2) > 0 $$
for $u_1 > u_2$, and inequalities (\ref{l1>}) and (\ref{l2<}),
proves 
$${\pd q(b, u_2, a) \over \pd u_2} < 0 $$
for $u_2 < u^{***}$. 
Hence,
$${\pd \mu_2 \over \pd u_2} = {1 \over 2} {\pd \lambda_2 \over \pd u_2}{\pd q \over \pd u_2}
+ {1 \over 2} [\lambda_2 - 2(b + u_2 + a)] {\pd^2 q \over \pd u_2^2} < 0 \ .$$

\end{proof}

\subsection{Type VII: $a < 0$, $b = -2a$}

\begin{figure}[h] \label{fig7}
\begin{center}
\resizebox{12cm}{4cm}{\includegraphics{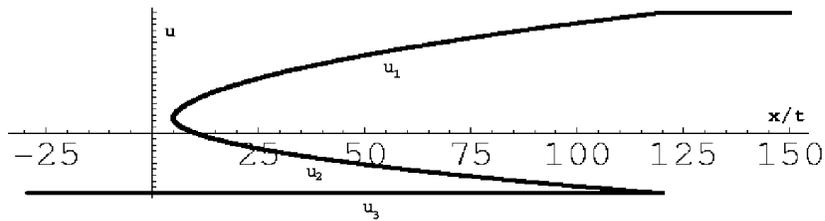}}
\caption{Self-Similar solution of the Whitham equations for
  $a=-1$ and $b= 2$ of type VII.}
\end{center}\end{figure}

\begin{thm}(see Figure 7.)
For the step like initial data (\ref{step}) with $a < 0$, $b = -2a$,
the solution of the Whitham equations
(\ref{5KdVW}) is given by
\begin{equation*}
x = \mu_1(u_1, u_2, a) \ t \ , \quad x = \mu_2(u_1, u_2, a) \ t \ , u_3 = a
\end{equation*}
for $5 a^2 < x/t < 120 a^2$.
Outside the region, the solution of the Burgers
type equation (\ref{5Burgers}) is given by
\begin{equation*}
u \equiv a \quad \mbox{$x/t \leq  5 a^2$} 
\end{equation*}
and
\begin{equation*}
u \equiv b \quad \mbox{$x/t \geq 120 a^2$} \ .
\end{equation*}
\end{thm}

\begin{proof}

It suffices to show that $u_1$ and $u_2$ of $\mu_1(u_1, u_2, a) - \mu_2(u_1, u_2, a)=0$
reaches $b = -2a$ and $a$, respectively, simultaneously. To see this, we deduce from
equation (\ref{2}) that
\begin{equation}
\label{7} \mu_1(u_1,a,a) - \mu_3(u_1,a,a) = 8 (u_1 -a)(u_1 + 2a)
\end{equation}
is negative for $u_1 < b$ and vanish when $u_1=b=-2a$.

\end{proof}

\subsection{Type VIII: $a < 0$, $b > -2a$}

\begin{thm}(see Figure 8.)
For the step like initial data (\ref{step}) with $a < 0$, $b > -2a$,
the solution of the Whitham equations
(\ref{5KdVW}) is given by
\begin{equation*}
x = \mu_1(u_1,u_2,a) \ t  \ , \quad x = \mu_2(u_1,u_2,a) \ t \ , \quad u_3 = a
\end{equation*}
for $5 a^2 < x/t < 120 a^2 $.
Outside the region, the solution of the Burgers
type equation (\ref{5Burgers}) is given by
\begin{equation*}
u \equiv a \quad \mbox{$x/t \leq 5 a^2$} 
\end{equation*}
and
\begin{equation*}
u = \left\{ \begin{matrix} \sqrt{{x \over 30 t}}
& 120 a^2 \leq x/t \leq 30b^2 \\
b & x/t \geq 30b^2 \end{matrix} \right. \ .
\end{equation*}
\end{thm}

\begin{proof}

By the calculation (\ref{7}), when $u_2$ of $\mu_2(u_1, u_2, a) - \mu_3(u_1, u_2, a)=0$
touches $a$, the corresponding $u_3$ reaches $-2a$, which is below $b$. Hence, equations
$$x = \mu_2(a, u_2, u_3) \ t \ , \quad x = \mu_3(a, u_2, u_3) \ t$$
can be inverted to give $u_2$ and $u_3$ as functions of $x/t$ in the region
$\mu_2(-a/4, -a/4, a) < x/t < \mu_2(-2a,-2a, a)$.  To the right of this region,
the Burgers type equation (\ref{5Burgers}) has a rarefaction wave solution.

\end{proof}

\begin{figure}[h] \label{fig8}
\begin{center}
\resizebox{12cm}{4cm}{\includegraphics{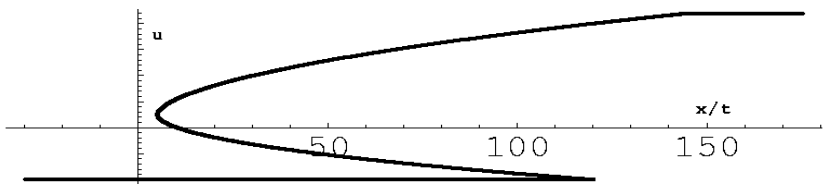}}
\caption{Self-Similar solution of the Whitham equations for
  $a=-1$ and $b= 2.5$ of type VIII.}
\end{center}\end{figure}

{\bf Acknowledgments.} We would like to thank Yuji Kodama and David Levermore for 
valuable discussions.
V.P. was supported in part by NSF Grant DMS-0135308. F.-R. T. was supported in part by 
NSF Grant DMS-0404931.

\bibliographystyle{amsplain}

\begin{thebibliography}{7}

\bibitem{dub} B.A. Dubrovin and S.P. Novikov, ``Hydrodynamics of Weakly
Deformed Soliton Lattices. Differential Geometry and Hamiltonian Theory'',
Russian Math. Surveys 44:6 (1989), 35-124.


\bibitem{gur} A.V. Gurevich and L.P. Pitaevskii, ``Non-stationary Structure
of a Collisionless Shock Wave'', Soviet Phys. JETP 38 (1974), 291-297.

\bibitem{kri} I.M. Krichever, ``The Method of Averaging for
Two-dimensional `Integrable' Equations'', Functional Anal. App.
22 (1988), 200-213.

\bibitem{lax} P.D. Lax and C.D. Levermore, ``The Small Dispersion
Limit for the Korteweg-de Vries Equation I, II, and III'', Comm. Pure Appl.
Math.
36 (1983), 253-290, 571-593, 809-830.

\bibitem{lax2} P.D. Lax, C.D. Levermore and S. Venakides, ``The Generation and
Propagation of Oscillations in Dispersive IVPs and Their Limiting Behavior''
in Important Developments in Soliton Theory 1980-1990, T. Fokas and V.E.
Zakharov eds., Springer-Verlag, Berlin (1992).

\bibitem{lef} P.G. LeFloch, {\em Hyperbolic Systems of Conservation Laws},
Lectures in Mathematics, Birkhauser, 2002.

\bibitem{lev} C.D. Levermore, ``The Hyperbolic Nature of the Zero Dispersion
KdV Limit'', Comm. P.D.E. 13 (1988), 495-514.

\bibitem{Tian1} F.R. Tian, ``Oscillations of the Zero Dispersion Limit of the
Korteweg-de Vries Equation'', Comm. Pure Appl. Math. 46 (1993), 1093-1129.

\bibitem{Tian2} F.R. Tian, ``On the Initial Value Problem of the Whitham
Averaged
System'', in Singular Limits of Dispersive Waves, N. Ercolani, I. Gabitov,
D. Levermore and D. Serre eds., NATO ARW series, Series B: Physics Vol. 320,
Plenum, New York (1994), 135-141.

\bibitem{Tian3} F.R. Tian, ``The Whitham Type Equations and Linear
Overdetermined
Systems of Euler-Poisson-Darboux Type'', Duke Math. Jour. 74 (1994), 203-221.

\bibitem{tsa} S.P. Tsarev, ``Poisson Brackets and One-dimensional
Hamiltonian Systems of Hydrodynamic Type'', Soviet Math. Dokl.
31 (1985), 488-491.

\bibitem{ven} S. Venakides, ``The Zero Dispersion Limit of the KdV Equation
with Nontrivial Reflection Coefficient'', Comm. Pure Appl. Math.
38 (1985), 125-155.


\bibitem{whi} G.B. Whitham, ``Non-linear Dispersive Waves'', Proc. Royal Soc.
London Ser. A 139 (1965), 283-291.


\end{thebibliography}

\end{document}